\documentclass{pasj00}
\usepackage{color}
\draft

\begin{document}
\SetRunningHead{Onaka et al.}{IRC for {\it AKARI}}
\Received{2007/03/02}
\Accepted{2001/04/09}

\title{The Infrared Camera (IRC) for {\it AKARI} - Design
and Imaging Performance}

\author{Takashi \textsc{Onaka},\altaffilmark{1}
        Hideo \textsc{Matsuhara},\altaffilmark{2}
        Takehiko \textsc{Wada},\altaffilmark{2}
        Naofumi \textsc{Fujishiro},\altaffilmark{3}\thanks{
        Present Address is Cybernet systems Co. Ltd., Bunkyo-ku, 
        Tokyo 112-0012, Japan}\\
        Hideaki \textsc{Fujiwara},\altaffilmark{1}
        Miho \textsc{Ishigaki},\altaffilmark{4}
        Daisuke \textsc{Ishihara},\altaffilmark{1}
        Yoshifusa \textsc{Ita},\altaffilmark{2}\\
        Hirokazu \textsc{Kataza},\altaffilmark{2}
        Woojung \textsc{Kim},\altaffilmark{2}
        Toshio \textsc{Matsumoto},\altaffilmark{2}
        Hiroshi \textsc{Murakami},\altaffilmark{2}\\
        Youichi \textsc{Ohyama},\altaffilmark{2}
        Shinki \textsc{Oyabu},\altaffilmark{2}
        Itsuki \textsc{Sakon},\altaffilmark{1}
        Toshihiko \textsc{Tanab\'e},\altaffilmark{5}\\
        Toshinobu \textsc{Takagi},\altaffilmark{2}
        Kazunori \textsc{Uemizu},\altaffilmark{2}
        Munetaka \textsc{Ueno},\altaffilmark{6}
        Fumio \textsc{Usui},\altaffilmark{2}
        Hidenori \textsc{Watarai},\altaffilmark{7}\\
        Martin \textsc{Cohen},\altaffilmark{8} 
        Keigo \textsc{Enya},\altaffilmark{2}  
        Takafumi \textsc{Ootsubo},\altaffilmark{9}
        Chris P. \textsc{Pearson},\altaffilmark{2,10}
\and       
        Norihide \textsc{Takeyama},\altaffilmark{11}
        Tomoyasu \textsc{Yamamuro},\altaffilmark{11}\thanks{
        Present address: OptCraft, Hadano, Kanagawa, 259-1331, Japan}
        Yuji \textsc{Ikeda},\altaffilmark{11}\thanks{
        Present address: Photocoding, Higashi-Hashimoto,
	Sagamihara, Kanagawa, 229-1104, Japan}
        }

\altaffiltext{1}{Department of Astronomy, Graduate School of Science,
The University of Tokyo, \\
Bunkyo-ku, Tokyo 113-0033, Japan}
\email{onaka@astron.s.u-tokyo.ac.jp}
\altaffiltext{2}{Institute of Space and Astronautical Science, \\
Japan
Aerospace Exploration Agency, Sagamihara, Kanagawa 229-8510, Japan}
\altaffiltext{3}{Department of Physics, Graduate School of Science,
The University of Tokyo, \\
Bunkyo-ku, Tokyo 113-0033, Japan}
\altaffiltext{4}{Department of Physics, Faculty of Science,
Tokyo Institute of Technology, \\
Meguro-ku, Tokyo 15208551, Japan}
\altaffiltext{5}{Institute of Astronomy, Graduate School of Science,
The University of Tokyo, \\
Mitaka, Tokyo 181-0015, Japan}
\altaffiltext{6}{Department of Earth Science and Astronomy, 
Graduate School of Arts and Sciences,\\
The University of Tokyo, 
Meguro-ku, Tokyo 153-8902, Japan}
\altaffiltext{7}{Office of Space Applications, Japan Aerospace Exploration
Agency, \\
Tsukuba, Ibaraki 305-8505, Japan}
\altaffiltext{8}{Radio Astronomy Laboratory, 601 Campbell Hall, 
University of California,\\  
Berkeley, CA94720, U.S.A.}
\altaffiltext{9}{Graduate School of Science, Nagoya University, Furo-cho, 
Chikusa-ku, Nagoya 464-8602, Japan}
\altaffiltext{10}{ISO Data Center, European Space Agency,
Villafranca del Castillo, P.B.Box 50727,\\
28080 Madrid, Spain}
\altaffiltext{11}{Genesia Corporation, Shimo-renjaku,
Mitaka, Tokyo 181-0013, Japan}


%


\maketitle

\begin{abstract}

The Infrared Camera (IRC) is one of two focal-plane instruments on the 
{\it AKARI}
satellite.  It is designed for  
wide-field deep imaging and low-resolution spectroscopy 
in the near- to mid-infrared (1.8--26.5\,$\mu$m) in
the pointed observation mode of {\it AKARI}.  IRC is also operated in
the survey mode to make an all-sky survey at 9 and 18\,$\mu$m. 
It comprises three channels.  The NIR channel (1.8--5.5\,$\mu$m) employs
a $512 \times 412$ InSb array, whereas both the MIR-S (4.6--13.4\,$\mu$m) and
MIR-L (12.6--26.5\,$\mu$m) channels use $256 \times 256$ Si:As impurity band
conduction arrays.  Each of the three channels has a field-of-view of 
about $10^\prime \times 10^\prime$ and are operated simultaneously.  
The NIR and MIR-S
share the same field-of-view by virtue of a beam splitter.  The MIR-L
observes the sky about $25^\prime$ away from the NIR/MIR-S field-of-view.
IRC gives us deep insights into the formation and evolution of galaxies,
the evolution of planetary disks, the process of star-formation, the 
properties of interstellar matter under various physical conditions, and
the nature and evolution of solar system objects.  The
in-flight performance of IRC has been confirmed
to be in agreement with the pre-flight expectation.  This paper summarizes
the design and the in-flight operation and imaging 
performance of IRC.

\end{abstract}

\section{Introduction}

The {\it AKARI}
spacecraft may be fixed in inertial space for up to approximately
600s to make a pointed observation in addition to the continuous
survey mode, in which all-sky survey observations
are performed \citep{Murakami07}.  
The Infrared Camera (IRC) on board {\it AKARI} is uniquely designed for
deep imaging and spectroscopy for pointed observations,
while it is also operated in the survey mode \citep{Onaka04}.

Because of the nature of near-earth orbits,
severe visibility constraints are imposed on pointed observations of {\it AKARI}
\citep{Murakami07}.  The IRC is designed to maximize the observation
efficiency in a pointed observation.
It consists of three channels
divided by the spectral range, each of which has a wide field-of-view
of about $10^\prime \times 10^\prime$ at the cost of the spatial resolution.
The IRC is designed as a general-purpose instrument to obtain 
information on the spectral energy
distribution of an object in the near- to mid-infrared.  
The imaging bands are selected to evenly cover the
spectral range (1.8 -- 26.5\,$\mu$m) limited by the detector responses 
as much as possible.  
The IRC has a unique feature of slit-less spectroscopy to obtain
spectra of multiple objects in a pointed observation for the first time
in space-borne instruments \citep{Ohyama07}.
The worth of imaging and low-resolution spectroscopy in the near- to 
mid-infrared has well been demonstrated by successful observations of
ISOCAM on board the Infrared Space Observatory \citep{Cesarsky96}.  IRC
has a much wider field-of-view and a longer wavelength coverage with
fewer filter selections than the ISOCAM.
Compared to the Infrared Array Camera (IRAC) 
on {\it Spitzer} \citep{Fazio04}, IRC has a wider field-of-view
with larger pixel scales
and spectral coverage longer than
10\,$\mu$m.  The IRC also has a spectroscopic capability for wavelengths
shorter than 5\,$\mu$m, which the Infrared Spectrograph (IRS) on {\it Spitzer} 
does not cover \citep{Houck04}.   Small slits in the IRC
enable spectroscopy for diffuse sources.
IRC targets range from distant galaxies
to solar system objects.  The prime scientific themes include the formation and 
evolution of
galaxies, the evolution of planetary disks, the process of star-formation,
the properties of interstellar medium, and the nature of solar system
objects, by means of deep imaging and spectroscopic surveys of wide
areas of the sky \citep{Matsuhara05, Matsuhara06}.  

This paper reports the
basic design and the in-flight performance of IRC imaging observations.
The performance in the spectroscopic mode is
given separately in \citet{Ohyama07}.  
The data reduction of the all-sky observation mode is described in 
\citet{Ishihara07}.
The in-flight calibration as well as
the stability of the performance will be given in Tanab\'e et al. (in prep.)

\section{Instrumentation Description}

\begin{figure}
  \begin{center}
    \FigureFile(120mm,120mm){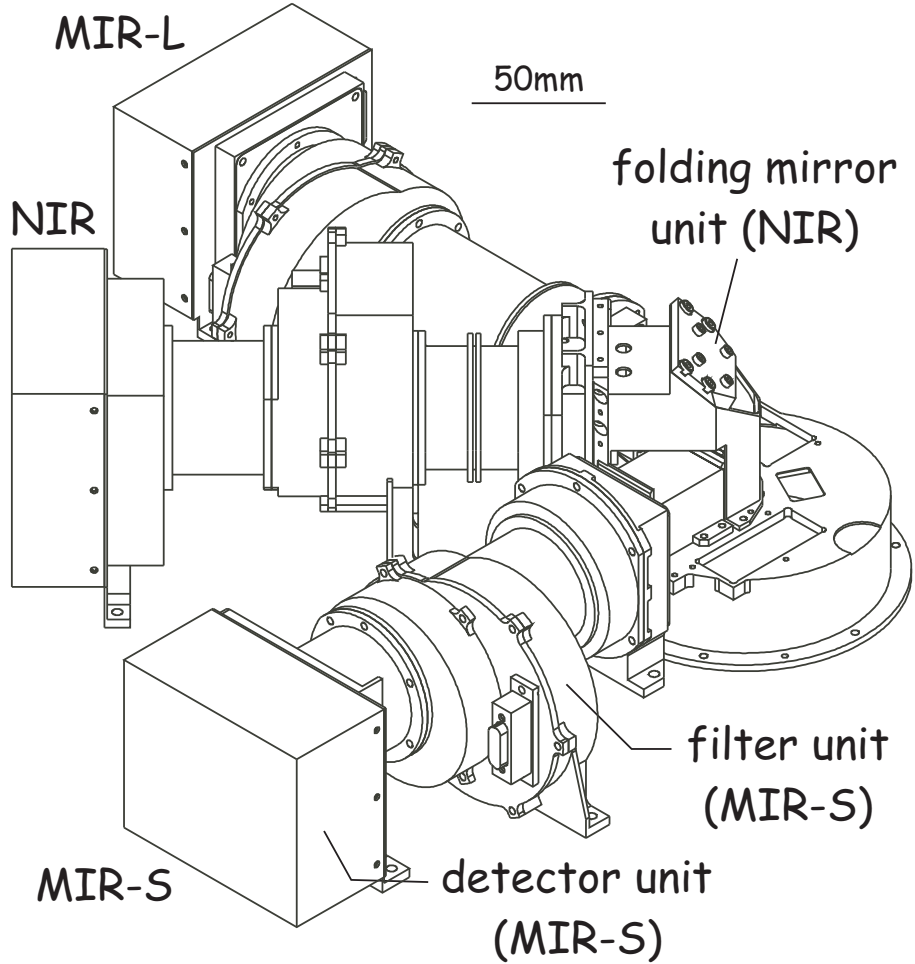}
  \end{center}
  \caption{Bird's-eye view of the IRC.}\label{figIRC}
\end{figure}

\begin{figure}
  \begin{center}
    \FigureFile(140mm,120mm){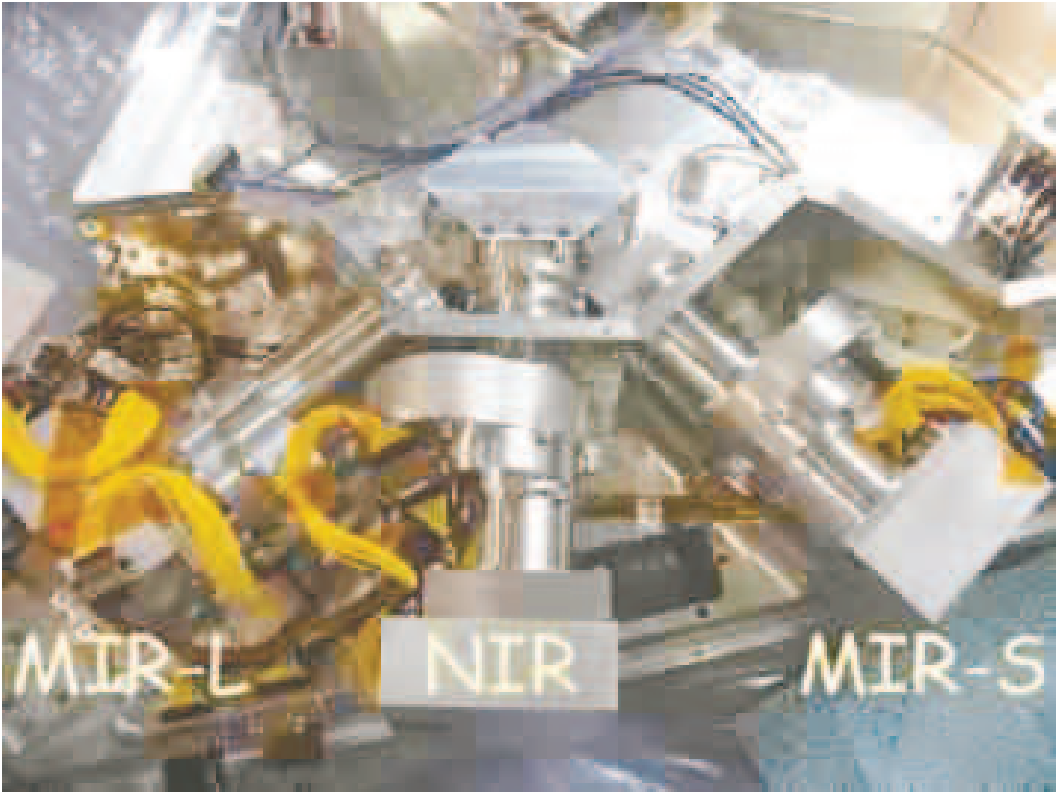}
  \end{center}
  \caption{Photograph of the IRC on the FPI plate with
  the telescope behind the plate.  From left to right, MIR-L, NIR, and
  MIR-L.}\label{figphoto}
\end{figure}

IRC consists of three channels:
the NIR channel 
operates from 1.8--5.3\,$\mu$m; the MIR-S channel works in 5.4--13.1\,$\mu$m; 
and the MIR-L channel
covers 12.4--26.5\,$\mu$m.  All three channels are operated simultaneously.
The NIR employs a $512 \times 412$ InSb
array, whereas the MIR-S and MIR-L use $256 \times 256$ Si:As arrays.
The InSb array operates at about 10K and the Si:As arrays at 6.5--7K.  To 
achieve
these temperatures in a stable condition, the InSb array unit has a weak thermal
connection to the focal-plane instrument (FPI) plate, whereas the
MIR detector units are connected firmly to the plate.
The FPI plate is weakly connected to the helium tank and is cooled by
the evaporating helium gas to about 6\,K \citep{Nakagawa07}.
The IRC arrays are kept running all the time and the temperatures
of the arrays are kept stable by the balance between their self-heating
and the cooling to the FPI plate.  The heat dissipation 
is approximately 0.6\,mW per array.
Both arrays were supplied by Raytheon, California, U.S.A.  The Si:As arrays
are of the same type as those employed by IRAC on {\it Spitzer} \citep{Fazio04}, 
while the InSb array is a larger format version of those on IRAC.  All 
three channels use refractive optics to enable a compact design
and achieve a nearly diffraction-limited performance.   
The final image quality is
dominated by the wave-front errors of the telescope for the NIR channel
\citep{Kaneda07}.  Each channel is equipped with three filters for
imaging observations,
two dispersive elements (prism and grisms)
for spectroscopic observations, and the shutter for the dark measurement,
all of which are installed in the filter wheel with a stepping motor
and each of which is selected by 
rotating the wheel.   The stepping motor was specially developed for
cryogenic applications by Sumitomo Heavy Industry.  The filter
wheel mechanism has extensively been tested on the ground and
has never failed so far.  It allows us to make multi-filter observations
and the measurement of dark current in one pointing.
The photometric band and disperser parameters are summarized
in Table~\ref{tableIRC}.  A bird's-eye view of the IRC is shown
in Figure~\ref{figIRC}.  Figure~\ref{figphoto} shows a photograph of
the IRC installed on the focal plane.  

\begin{table}
\caption{IRC band characteristics}\label{tableIRC}
\begin{center}
\begin{tabular}{cccccccc}
\hline
\hline
(1)     & (2)     & (3)    & (4)         & (5)    & (6)   & (7)  & (8)\\
Channel$^*$ & Name    & element & $\lambda_{\rm ref}$    & Wavelength  & 
$\lambda_{\rm iso}$ & $\Delta\lambda$ & Dispersion \\
(pixel scale)        &         &        & ($\mu$m)    & ($\mu$m) & ($\mu$m) & 
($\mu$m)  & ($\mu$m/pix) \\
\hline
        & N2      & filter & 2.4 & 1.9--2.8    &  2.34  & 0.71  & --- \\
NIR        & N3      & filter & 3.2 & 2.7--3.8    &  3.19  & 0.87  & --- \\
($1.^{\prime\prime}46 \times 1.^{\prime\prime}46$)    & N4      & filter & 4.1 & 
3.6--5.3    &  4.33  & 1.53 & --- \\
        & NP      & prism  &     & 1.8--5.5    &  ---   & ---   & 0.06 @3.5 
$\mu$m$^\dagger$ \\
        & NG      & grism  &     & 2.5--5.0    &  ---   & ---   & 0.0097  \\
\hline
        & S7      & filter & 7.0  & 5.9--8.4    &  7.12   & 1.75   & --- \\
MIR-S        & S9W     & filter & 9.0  & 6.7--11.6   &  8.61   & 4.10   & --- \\
($2.^{\prime\prime}34 \times 2.^{\prime\prime}34$)  & S11     & filter & 11.0 & 
8.5--13.1   & 10.45   & 4.12   & --- \\
        & SG1     & grism  &      & 4.6--9.2    & ---    & ---   & 0.057  \\
        & SG2     & grism  &      & 7.2--13.4   & ---    & ---   & 0.097 \\
\hline
        & L15     & filter & 15.0 & 12.6--19.4  & 15.58   & 5.98  &--- \\
MIR-L        & L18W    & filter & 18.0 & 13.9--25.6  & 18.39   & 9.97  &--- \\
($2.^{\prime\prime}51 \times 2.^{\prime\prime}39$)   & L24     & filter & 24.0 & 
20.3--26.5  & 22.89   & 5.34   &--- \\
        & LG2     & grism  &      & 17.5--26.5  & ---    & ---   &  0.17 \\
\hline

\multicolumn{8}{l}{(4) Reference wavelength.}\\
\multicolumn{8}{l}{(5) Defined as where the responsivity is 
larger than $1/e$ of the peak for the imaging mode.}\\
\multicolumn{8}{l}{
See \citet{Ohyama07} for the spectroscopic modes.}\\
\multicolumn{8}{l}{(6) Isophotal wavelength of the filter band for Vega.}\\
\multicolumn{8}{l}{(7) Effective bandwidth.}\\
\multicolumn{8}{l}{$^*$ All the channels have fields-of-view of about 
$10^\prime \times 10^\prime$ (see Fig.~\ref{figFP}).}\\
\multicolumn{8}{l}{$^\dagger$ Dispersion power of NP depends on the 
wavelength.}\\
\end{tabular}
\end{center}
\end{table}

\begin{figure}
  \begin{center}
    \FigureFile(160mm,120mm){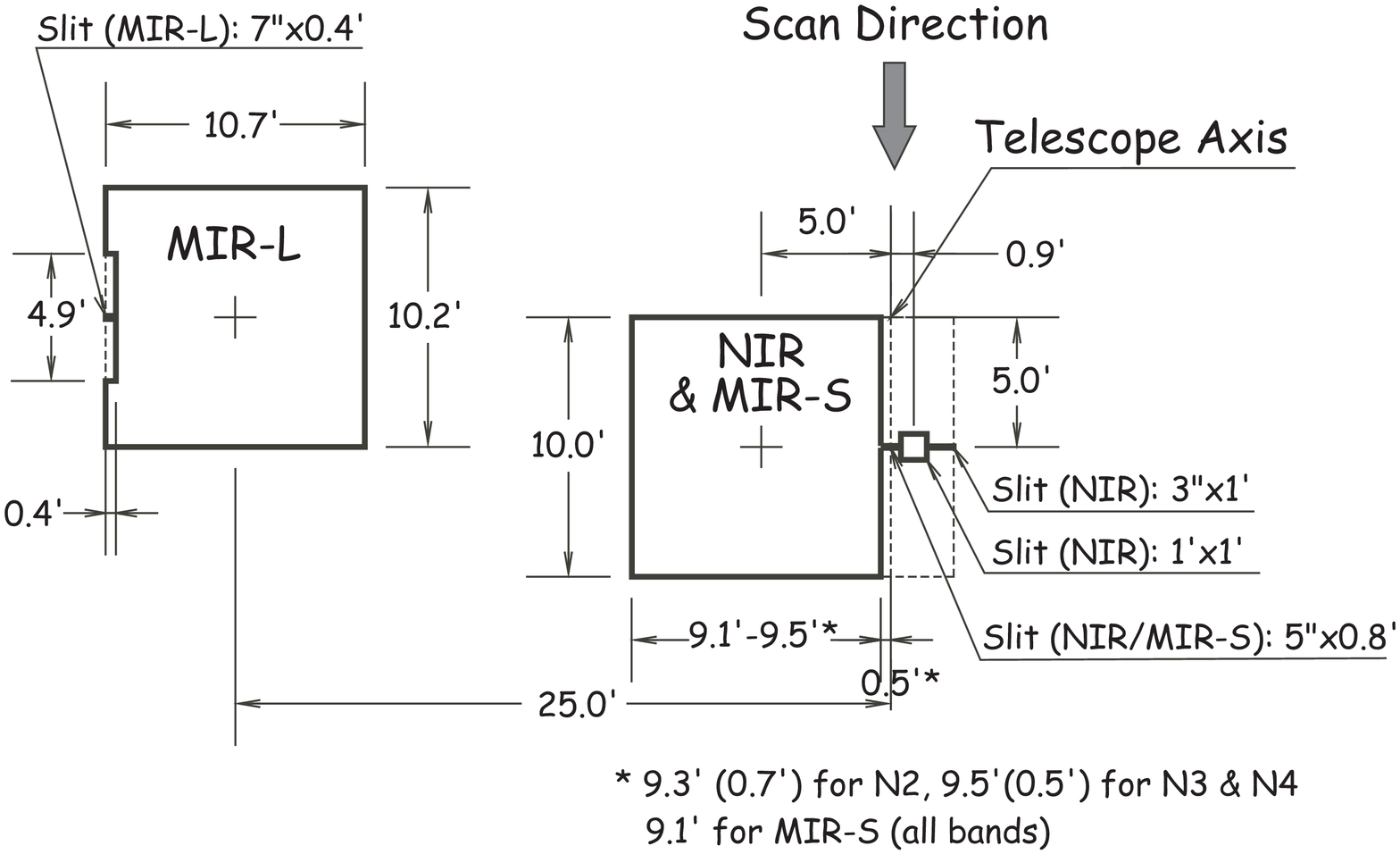}
  \end{center}
  \caption{Field-of-view location of the three IRC channels.
  The vertical arrow indicates the scan direction in the survey mode.
  The NIR and MIR-S share the same field-of-view by means of the
  beam splitter.  See text for the NIR slit usage.}\label{figFP}
\end{figure}

\begin{figure}
  \begin{center}
    \FigureFile(115mm,100mm){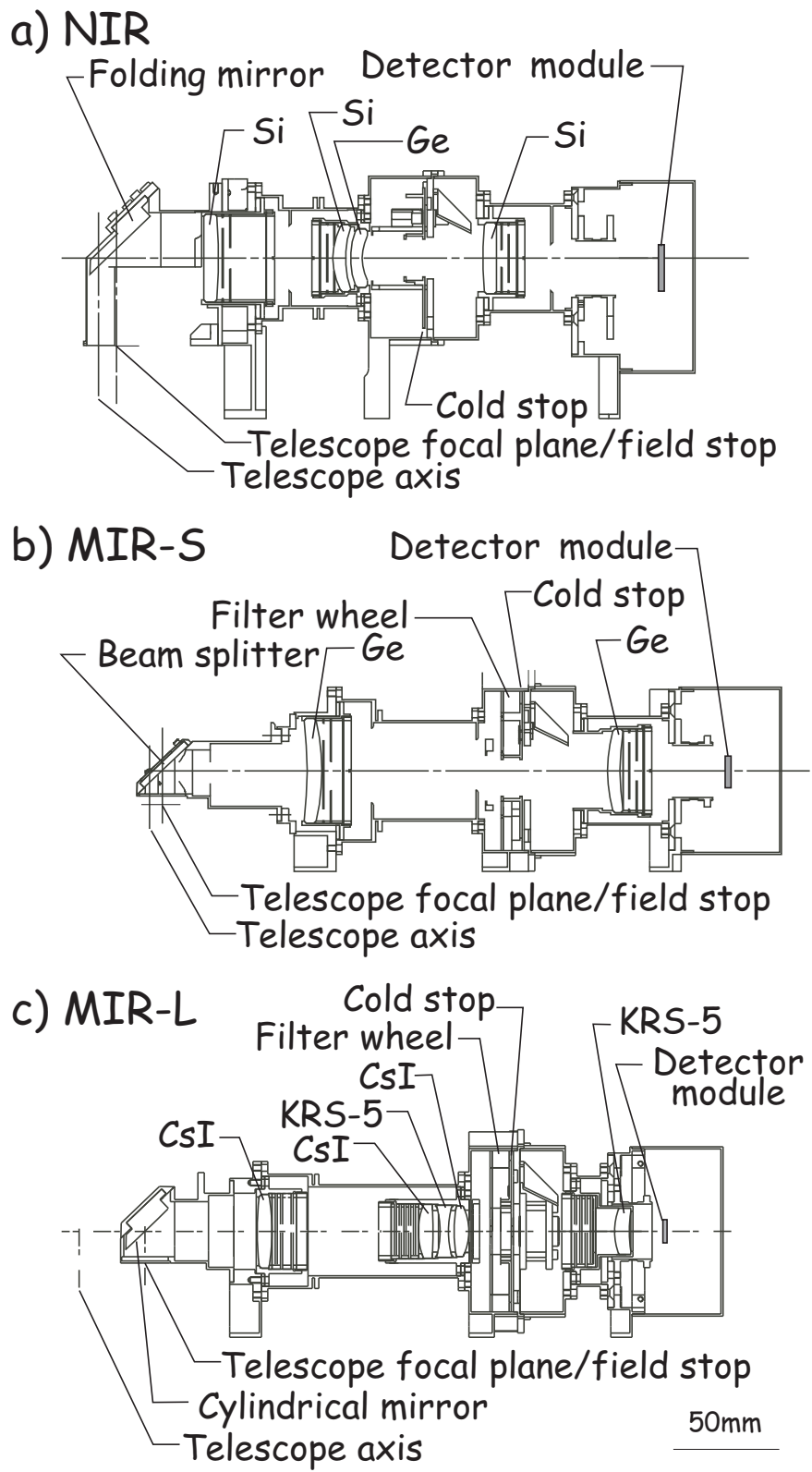}
  \end{center}
  \caption{Side-view of the IRC three channels. a) NIR, b) MIR-S, and c)
  MIR-L.}\label{figopt}
\end{figure}

Figure~\ref{figFP} shows the layout of the fields-of-view of each IRC channel.
The NIR and MIR-S share the same field-of-view by virtue of the 
beam splitter.
The MIR-L has a field-of-view at $25^\prime$ away from that of the NIR/MIR-S,
taking account of the focal-plane curvature.  The dispersers are all of 
transmission type
and thus IRC spectroscopic observations can be carried out in a
slit-less mode.  In addition, every channel has a small slit area as shown in 
Figure~\ref{figFP} and slit-spectroscopy can also be performed.  These
slits are installed in the field stops at the telescope 
focal plane as shown in figure~\ref{figopt}.
The NIR detector array has an extra part of
$100 \times 412$ pixels in addition to the square area, which provides an
extra slit area.  The NIR slit consists of a wide part in the middle
($1^\prime \times 1^\prime$), which is used for spectroscopy of point sources
with the grism (NG) to avoid confusion with nearby sources,
and the narrow parts at the left and right 
are used for spectroscopy of
diffuse sources.  The left 
part ($5^{\prime\prime} \times 0.4^\prime$)
is shared with the MIR-S.  
The small slit of the MIR-L is located at the opposite side 
of the
field-of-view compared to the NIR and MIR-S channels 
and is also designed for spectroscopy of diffuse sources.  
The absolute accuracy of
the telescope pointing is generally better than 3$^{\prime\prime}$.  
This is
sufficient to place a point source in the large 
$1^\prime \times 1^\prime$ slit for NIR.
Details of the mechanical design are given in
\citet{Watarai00}.  IRC weighs approximately 3.9kg in total, 
including the focal-plane interface nicknamed ``Swiss cheese'', which has
many holes for the field stops.

\begin{figure}
  \begin{center}
    \FigureFile(150mm,120mm){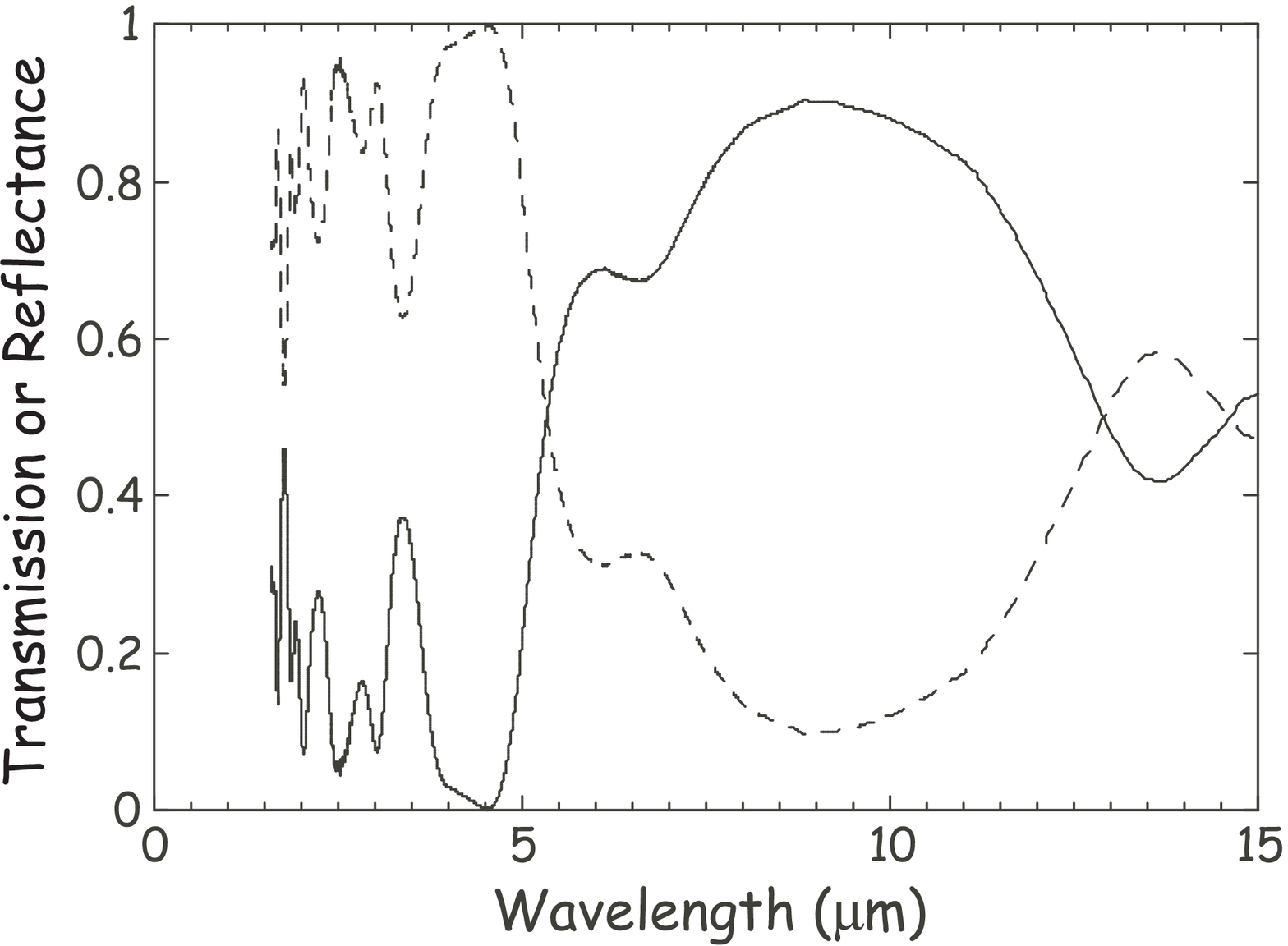}
  \end{center}
  \caption{Transmission and reflectance of the beam splitter
  at 4.2\,K.  The dashed line indicates the transmission and
  the solid line the reflectance.  See text for details.
  \label{figbs}}
\end{figure}

Figure~\ref{figopt} depicts the optical design of the three IRC channels.
The light from the telescope is split by the germanium beam splitter and the
transmitted light is introduced to the NIR channel.  The reflected light
goes to the MIR-S. 
Figure~\ref{figbs} shows the transmission and reflectance of 
the beam splitter.
The transmission was measured at 4.2\,K in normal incidence.
The shift in the wavelength for the 45 degree incidence 
was estimated from the transmission measurement
at room temperature between the normal incidence and 
the 45 degree incidence.  The reflectance was simply
estimated from the transmission, which was confirmed by the 
reflectance measurement at room temperature.  As can be seen in the
figure, the transmission is better than
the requirement of 70\% across most of the spectral range 
below 12\,$\mu$m.  However
the beam splitter is not perfect in any sense and in fact produces 
ghost images by reflections within it particularly at wavelengths where
the transmission becomes worse ($> 12$\,$\mu$m, see below).  
The dip at 3.4\,$\mu$m is seen exactly at the same wavelength
in the system throughput of the NIR spectroscopic modes
\citep{Ohyama07}, providing an independent confirmation of 
the transmission curve of the beam splitter.  The cold
stop is located right behind the filter wheel as shown in figure~\ref{figopt}
for all three channels.
Details
of the optical and mechanical design of the NIR channel are given in 
\citet{Kim05}.  The NIR consists of silicon and germanium lenses.  
There are non-negligible
color aberrations between the N2 and the N3/N4 bands,
which were utilized for the telescope focus adjustment \citep{Kaneda07}.
They are smaller than the wave-front errors of the telescope,
which dominate in the total wave-front errors for the NIR channel.
The NIR has a prism consisting of silicon and CaF$_2$ and 
a germanium grism as the dispersers.  The grism provides medium
resolution spectroscopic capability in the near-infrared.
The pixel scale of the NIR channel is 1.$^{\prime\prime}$46.

The MIR-S channel optics consist of two aspherical germanium lenses.
It has two medium band filters (S7 and S11) and a wide-band filter (S9W),
the latter of which is used for deep survey as well as for the all-sky survey 
observations.  It also has two grisms of KRS-5 to cover the wavelength range
4.6 to 13.4\,$\mu$m.  The pixel scale of the MIR-S channel is 
2.$^{\prime\prime}$34.

The MIR-L has the most complicated optical design: It has a cylindrical
pick-up mirror to correct for astigmatism.  The MIR-L channel consists of
3 CsI and 2 KRS-5 lenses.  Because of the cylindrical
mirror, the pixel scale is different in the orthogonal directions,
being $2.^{\prime\prime}51 \times 2.^{\prime\prime}39$.  The scale
in the horizontal direction of Figure~\ref{figFP} is longer than
that in the vertical direction.
Details of the MIR-L design are given
in \citet{Fujishiro05}.  The MIR-L has two medium-band filters
(L15 and L24) and a wide-band filter (L18W), the latter of which is
used for the deep survey as well as the all-sky survey observations.  
The MIR-L had
two grisms of KRS-5, but one of them (LG1), which was designed to
cover the 11--19\,$\mu$m spectral range, became
opaque in the last stage of ground testing and the spectroscopic
capability between 13 -- 18\,$\mu$m was lost.  Part of the edges of the
spectral range of LG1
are covered by SG1 and LG2 (see Table~\ref{tableIRC}).

The warm electronics of IRC is installed in the same electronics box 
as the other
on-board scientific instrument, the Far-Infrared Surveyor (FIS; 
\cite{Kawada07}).  The electronics box is thermally insulated
and attached to the outer shell of the cryostat \citep{Nakagawa07}.
It controls the array operation, takes signals, and sends them
to the Data Handling Unit (DHU) of the spacecraft.  It also receives the
signal of the attitude stability for the pointing observation, which 
initiates the array operation sequence.  
The warm electronics of IRC consists of
the power supply, the array drivers, the peripheral control, the sequencer,
and the on-board computer unit (CPU) boards.  The clock pattern for the
array operation is sent to the sequencer by a trigger from the CPU.
The array operation is carried out by the sequencer independently and
thus it is not affected by the CPU status.  Details of the IRC electronics
and the array operations
are given in \citet{Wada03} and \citet{Ishihara06a}.

\section{Focal-plane array operation and observation mode}
 
In the pointing mode, a combination of the exposure, filter exchange, and
dithering operations is fixed to several patterns, which are
called Astronomical Observation Template (AOT).  The IRC AOT is designated
as IRCxx, where xx indicates the AOT number.  The IRC AOT in use for
actual astronomical observations
includes one-filter mode (IRC05), two-filter mode (IRC02), three-filter mode 
(IRC03), 
spectroscopic mode (IRC04), and slow-scan mode (IRC11 and IRC51).
IRC05 was chosen on orbit to be used as the one-filter mode, for which
IRC00 was originally planned.  Since IRC00 has rarely been used on orbit,
only IRC05 is described in the present paper. 
In each AOT, there are additional parameters for the selection of filters, 
dispersers, field-of-view location, and usage of the slit.  

\begin{figure}
  \begin{center}
    \FigureFile(160mm,120mm){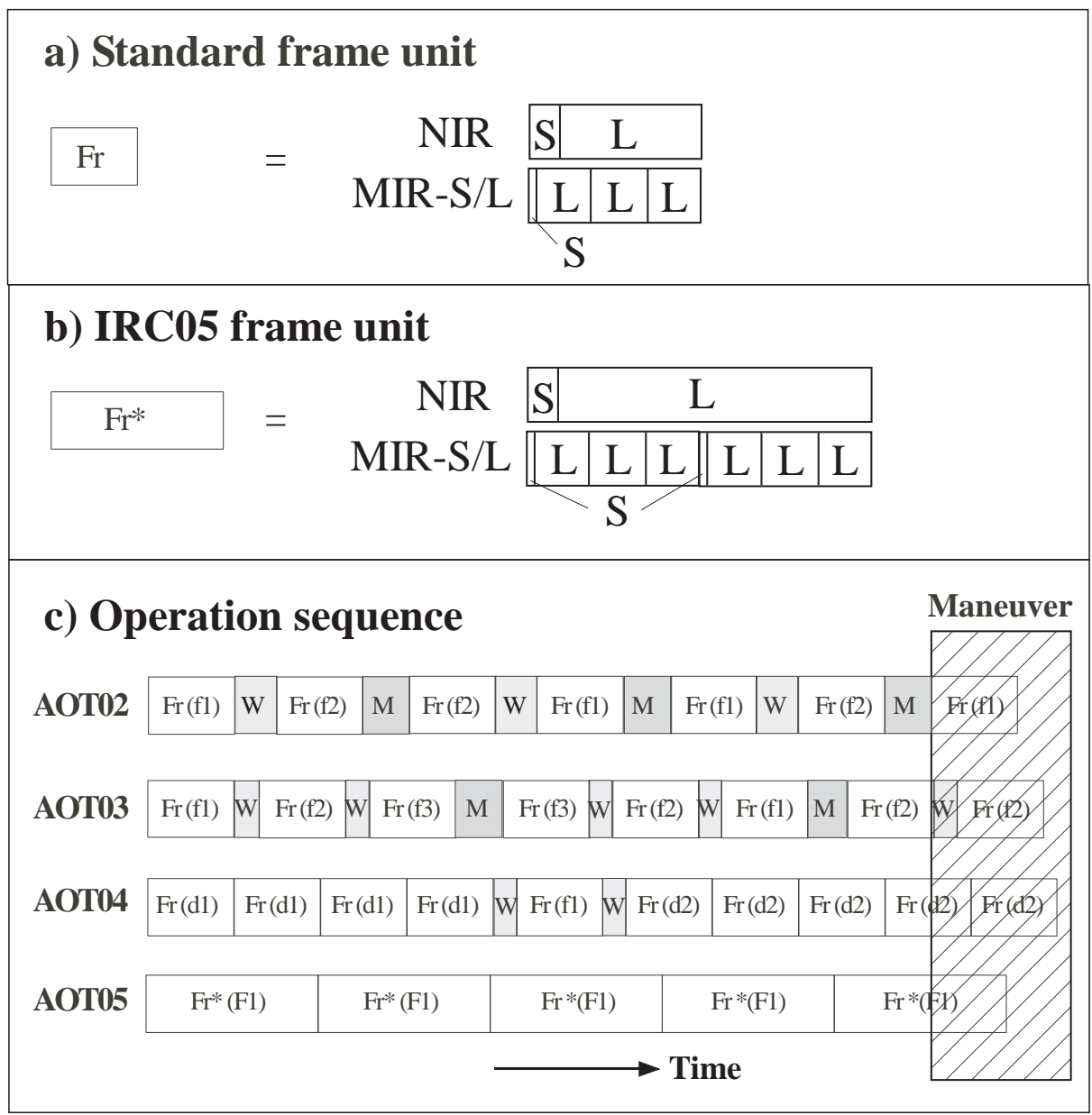}
  \end{center}
  \caption{
Unit frame and
operation sequence of the major IRC AOTs used on orbit.
(a) Standard frame unit ($Fr$) and (b) IRC05 frame unit ($Fr^*$); 
$L$ indicates the long exposure and $S$ shows the short exposure.  
(c) Operation sequences for four IRC AOTs.
The boxes with
$Fr$ and $Fr^*$ show standard frames and IRC05 frames, respectively,
in which f\# and d\# in the parentheses 
indicate the filter and dispersive element, respectively.
The boxes with $M$
indicate a dithering operation, whereas $W$ shows a filter wheel operation.
The shaded boxed in the right indicates the region, in which the satellite
maneuver may be carried out and thus the frames may be taken while
the target is moving.  Whereas the duration of one exposure frame
is well fixed, the operation time for the filter wheel and
dithering depends on the relative positions of the wheel and the attitude
control system.  Hence the length of the box is not very accurate.
  \label{figAOT}}
\end{figure}

A unit of the exposure pattern of the focal-plane array is 
the same for all the AOTs
except for IRC05.  
The unit is called `exposure frame'.  The unit frame except for
IRC05 is called a standard frame.
The unit exposure frame for IRC05 is defined as an IRC05 frame and is
described separately.  The operation of the
MIR-S and MIR-L arrays is controlled by the same clock and thus has the
same exposure pattern \citep{Wada03}. 
It takes about 70s to execute one standard frame. 
One standard frame consists of short
and long exposures as shown in figure~\ref{figAOT}a.  
The short exposure is accommodated to increase
the dynamic range: One standard frame of the NIR channel consists of one
short exposure and one long exposure and 
the short exposure time is $1/9.5$ of the long
exposure frame.  
One standard frame of the MIR-S and MIR-L channels has
one short exposure and three long exposures and the short exposure time is
$1/28$ of the long one.  Data of each
array are taken using a Fowler sampling method \citep{Fowler90}.
The Fowler sampling method consists of N non-destructive reads immediately
after the reset (pedestal reads) and another N non-destructive reads
near the end of the integration (signal reads).  The average of pedestal
reads is subtracted from the average signal reads on board and transmitted
to the ground. 
In the short
exposure, the data are taken with Fowler 1 sampling.  The Fowler 4
sampling is employed for the long exposure in the standard frame.
The IRC05 
frame time is twice as long as the standard frame  
to enable the Fowler 16 sampling for NIR deep observations with one filter
as shown by $Fr^*$ in figure~\ref{figAOT}b.
One IRC05 frame has two standard frames of MIR-S and MIR-L.

A set of frames is arranged to accommodate each AOT.
Including the time for
the filter exchange and dithering operations, one pointed observation has
8 to 9 frames for IRC02, IRC03, and IRC04, and 4 or 5 frames for IRC05. 
Figure~\ref{figAOT}c 
shows a schematic diagram of the operations in four IRC AOTs.  
For IRC02, dithering operation is executed every two frames with different
filters.  For IRC03, three frames with three different filters are taken,
followed by one dithering operation.  In the IRC04 sequence,
there is no dithering operation.  After four consecutive frames with the same
dispersive element, one reference image frame is taken, and then another set of frames with another dispersive element is taken.  
The reference image is used to 
derive the wavelength reference position (see \cite{Ohyama07} for details).
IRC05 does not have any dithering or filter change operations.
 
In a pointed
observation, there is also one exposure frame before 
and after each pointed observation of the target during the
satellite maneuver, in which the filter wheel is positioned
at the shutter (pre- and post-dark frame).  For IRC11 and IRC51, measurements 
with the shutter closed are also made during the maneuver. 
Except for
IRC02, more than two pointed observations are recommended to obtain
at least three images of science targets for reliability and rejection of
high-energy ionization particle hits particularly for NIR observations
since only one or two frames of the same filter/disperser are taken with
AOTs other than IRC02.

The slow scan AOTs, IRC11 and IRC51, are designed to observe a relatively wide
area ($\sim 1^\circ$ strip) in one pointed observation.  The satellite moves
continuously with a slow rate of a fixed pattern.  The same array operation
used for the all-sky survey mode is employed 
in these observations \citep{Ishihara06a}.  Only
the data for two rows of each of
the MIR-S and MIR-L arrays are taken because of the
limitations of the data downlink capacity and the array drive electronics.
In IRC11, the data of 4 pixels in the cross-scan direction are binned together
in the same manner as in the all-sky survey observation 
to reduce the rate of the data
produced (see \S~\ref{all-sky}), 
while no binning is made for
IRC51.  As long as the downlink capacity is allowed, IRC51 is employed for
IRC slow scan observations.  Details of the in-flight operation 
of IRC51 are given in \citet{Ishihara07}.

\section{In-flight performance}
IRC is working without any degradation on orbit and
the in-flight performance has been confirmed to be
within the pre-flight expectation.

\subsection{System throughput}
\begin{figure}
  \begin{center}
    \FigureFile(160mm,120mm){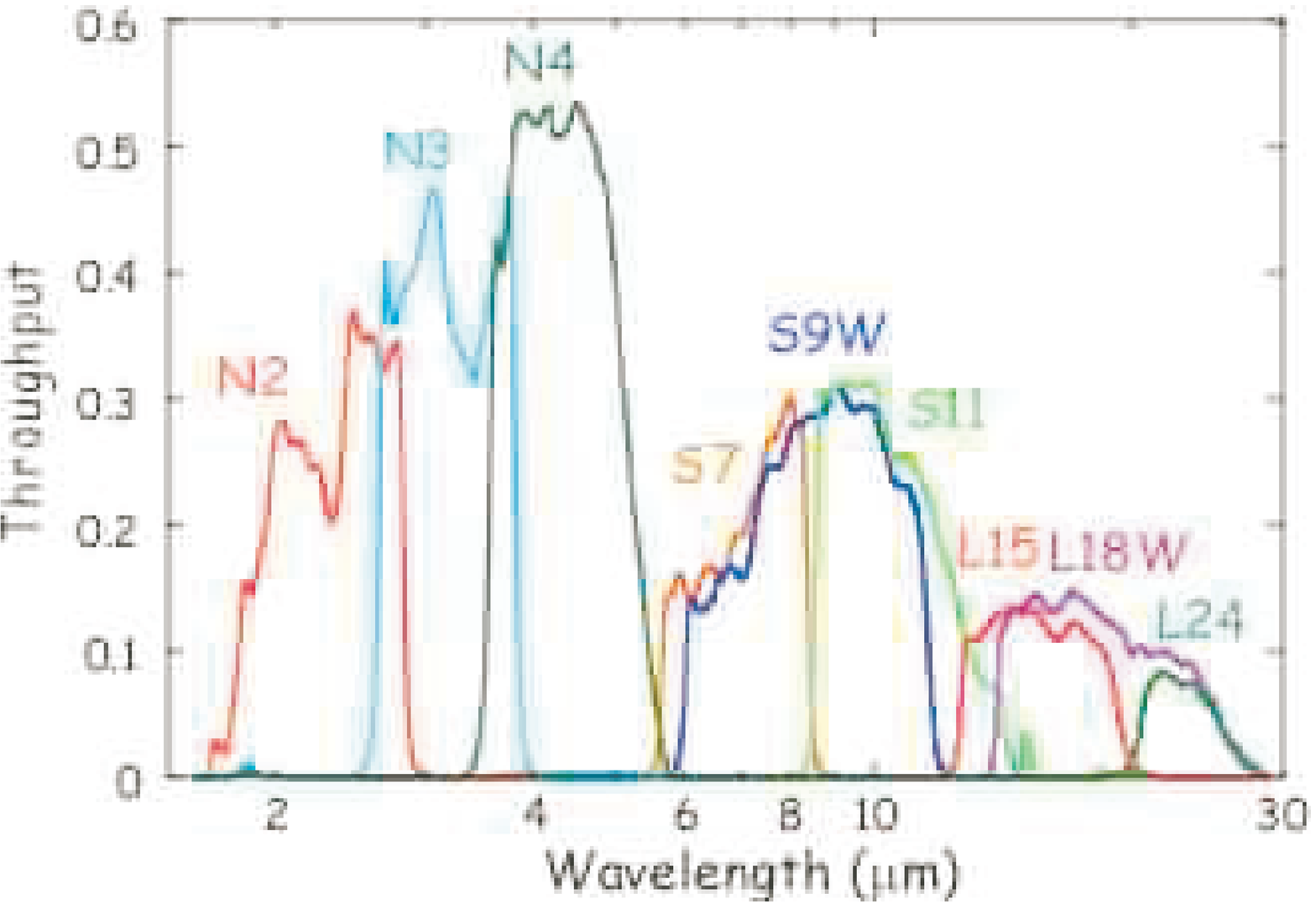}
  \end{center}
  \caption{System throughput of the IRC imaging bands
  (electrons per photon).}
  \label{figTPut}
\end{figure}

The spectral throughput in units of electrons per photon
of the imaging bands was estimated on the basis
of laboratory measurements of each optical element.  It is proportional
to the detector quantum efficiency.
It was confirmed
by end-to-end measurements of IRC without the telescope in the laboratory
except for the MIR-L, for which the end-to-end test was not able to be
carried out \citep{Onaka04}.  The final system throughput is scaled by
the in-flight absolute calibration and is shown in Figure~\ref{figTPut}.  
There are small red leaks in the 
N2 ($<$1.8\%) and N3 ($<$1\%) bands around 5\,$\mu$m.
Blue leaks in MIR-S and MIR-L shorter than 4 and 10\,$\mu$m, respectively, 
are recognizable, but they are all less than 0.01\%.
The system throughput for the spectroscopic mode is given in
\citet{Ohyama07}, which was derived from in-flight data.  It also basically 
agrees with pre-flight laboratory measurements.

\subsection{Calibration and stability}
The in-flight absolute calibration of IRC is made by observations
of standard stars prepared on the basis of the same standard
star networks as {\it MSX} and IRAC,
which assures 
that the IRC calibration is tied with IRAC and other facilities whose
calibration is based on the same networks \citep{Cohen99, Cohen03, Ishihara06b}.  
At present the accuracy of the absolute calibration for pointed observations
is estimated
to be better than 5\% except for L24, for which it is about 6\%.
The calibration of the flux density for a source is made at the reference 
wavelength $\lambda_{\rm ref}$
(4th column of Table~\ref{tableIRC}) for a nominal
spectrum of $\nu I_\nu =$ const.  A color correction table for
various spectra is given in the IRC Data User's  Manual \citep{IDUM}.

The stability of the system response has been 
investigated on the basis of monitoring
observations of the same 
stars in the continuous visibility regions every one to
two weeks.  No systematic trends have so far been seen over a half year. 
The system response is found to be stable within 5\%
for all the imaging modes.  The frequency of the monitoring observations will
be reduced owing to the excellent stability of IRC.

\subsection{Image quality and sensitivity}
The image size of the IRC is estimated on the basis of a number of stellar 
images.  Average values of the FWHM of the image size
are summarized in Table~\ref{tabSENSPSF}.
The image quality of the NIR channels is limited by the wave-front errors
of the telescope, but
part of the image quality of the NIR channels can also be attributed to
the color aberrations and the pointing stability of the satellite.
The image size in the L24 band is slightly larger than the predicted
performance.  The reason is not known at present.

The point-source sensitivity per one long exposure is estimated
from the aperture photometry of stars in low background regions.  
The aperture size is chosen to maximize the
signal-to-noise (S/N) ratio by taking account of the point spread function 
(PSF).
The sensitivity for each AOT in one pointed observation
is simply scaled from the estimate per long exposure.
The sensitivity of the NIR bands for IRC05 are 
estimated separately.  
The 5-$\sigma$ sensitivities per pointing
are summarized also in Table~\ref{tabSENSPSF}.
The effective on-source time is given in a footnote of
the table.  Also the saturation level for the short exposure is estimated
at the flux at which the error in the linearity correction becomes 5\%
at the central pixel (see next subsection) and is listed also in the table.
For MIR-S and MIR-L, the sensitivity is limited by shot noise from 
the sky background.  
The sensitivity on orbit is in agreement with the pre-flight
estimation within 20--30\%.

\begin{table}
\caption{IRC sensitivity and image quality}\label{tabSENSPSF}
\begin{center}
\begin{tabular}{ccccccc}
\hline
\hline
  & \multicolumn{3}{c}{5-$\sigma$ sensitivity$^*$}
  & Saturation$^\dagger$
& \multicolumn{2}{c}{Image quality} \\
Band &  \multicolumn{3}{c}{per pointing ($\mu$Jy)}& 
level &
FWHM  & Peak central \\
& AOT02 & AOT03 & AOT05  & (Jy) 
&(arcsec) &  pixel flux (\%) \\
\hline
N2  & 16  & 20   &  5$^\ddagger$  & 0.8 & 4.3  & 5.9 \\
N3  & 16  & 19   &  5$^\ddagger$  & 0.6 & 4.0  & 5.9 \\
N4  & 16  & 19   &  9$^\ddagger$  & 0.4 & 4.2  & 5.7 \\
S7  & 74  & 91   & 43  & 4.3 & 5.1  &12.0 \\
S9W & 76  & 93   & 44  & 2.7 & 5.5  &11.5 \\
S11 & 132 & 162  & 76  & 3.7 & 4.8  &12.1 \\
L15 & 279 & 341  & 161 & 9.9 & 5.7  & 7.9 \\
L18W & 273 & 335 & 158 & 8.4 & 5.7  & 6.6 \\
L24 & 584 & 716  & 337 & 70  & 6.8  & 4.1 \\
\hline
\multicolumn{7}{l}{$^*$ For low background.   
The effective on-source time per filter is
130, 90, 260s  }\\
\multicolumn{7}{l}{ for NIR and 150, 100, and 440s for
MIR-S and MIR-L, for AOT02, AOT03,  }\\
\multicolumn{7}{l}{ and AOT05, respectively. }\\
\multicolumn{7}{l}{ 
$^\dagger$ The saturation level for the short exposure (see text). }\\
\multicolumn{7}{l}{$^\ddagger$ The values for N2, N3, and N4 of IRC05 are 
tentative.} 
\end{tabular}
\end{center}
\end{table}

\subsection{Distortion and linearity}
The image distortion is estimated from observations of stars in
globular clusters except for L24
and those in a region near the Galactic center for L24.  At present the 
accuracy in the distortion
correction is about 0.1 pixel except for L24, for which it is
about half a pixel.
The worse value for L24 comes mainly from the small number statistics,
the scarcity of bright sources in a 
field-of-view of the MIR-L.

Because of the debias effects during an integration, an
appreciable
non-linearity exists for signals of all the IRC detectors.  
Polynomial fits up to the 
7th order
are used to correct for it.  After the correction, the error from the ideal
linearity is better than 5\% at 12000, 20000, and 20000 analog-to-digital 
units (ADUs) for NIR, MIR-S, and MIR-L, respectively.  
The physical detector saturation occurs at about 125000, 33000,
and 33000 ADUs for NIR, MIR-S, and MIR-L, respectively.

\subsection{Dark and flat field}

Pre- and post-dark frames before or after a pointed observation can be used
as the dark frame.  
Post-dark frames occasionally show
latent images from pointed observations and cautions have to be
given when they are used as the dark frame.
The pre-dark image does not have a high S/N
ratio.  It is particularly true for the NIR, for which the pre-dark data 
consist of only one long exposure. 
To have dark images of high S/N ratios,
`superdark' images were created from a number of pre-dark frames taken
in the early phase of the mission.  
The dark signals are found to vary after passage of the South Atlantic
Anomaly.  Thus the temporal variation must be corrected when the superdark
is used.  This can be done
by taking account of the offset between the superdark and the
dark signal before the target observation for the region masked by the
field stop.
The superdark images have the advantage of a high S/N ratio for
the pixel-to-pixel variation,
however, they do not correct for the increased dark signal of a band-like
pattern due to
the IRC all-sky survey observation in the MIR-S and MIR-L channels.
They also do not accommodate the increasing number of hot pixels.
Users have a choice which dark data are used for the data reduction.
The superdark is prepared mainly for the dark estimate for
NIR and for observations
whose pre-dark measurement failed. 

For the NIR bands, the flatfield library was made from observations
of the north ecliptic pole (NEP).  Since the sky is very faint in the 
near-infrared,
the S/N ratio is about 5 for N2 and N3, and 10 for N4 at present.

For the MIR channels, the flatfield library was made from observations
of the ecliptic plane (EP).  The S/N ratios for the MIR-S and MIR-L
bands are about 100.  The flat image of MIR-L shows a noticeable
pattern, which shows a large gradient ($\sim 30$\%) at the edges.  
It is surmised
that the flatfield for the MIR-L bands is 
affected by scattered light,
which presumably originates from internal reflections inside the MIR-L
channel.
Part of the flatfield pattern can be attributed to the scattered light.
The current estimate suggests that the flatfield for the MIR-L channels
may have an error of 10\% on a large spatial scale
due to the scattered light effect.

\subsection{Artifacts of the arrays}

The IRC arrays show similar artifacts to those known for the IRAC arrays,
such as multiplexer bleed, column pull-down, and banding \citep{Fazio04}.  
Details are described in the IRC Data User's Manual \citep{IDUM}.
So far, latent images do not
seem to be a serious problem probably because 
the arrays are switched to the all-sky survey mode and 
kept continuously running after the pointed
observation.  
The scattering inside the arrays (banding) is
under investigation and calibration for extended sources is being planned as
done by \citet{Cohen07} for IRAC.

\subsection{Ghosts and scattered light}\label{scat}
The NIR and MIR-S channels are known to have ghosts originating from
the reflection in the beam splitter.  They are most noticeable in the
S11 with a level of 4\%.  For other NIR and MIR-S bands,
they are less than 1\%.
These ghosts are well predictable in position and intensity
and thus can be corrected.

There are several kinds of known scattered light in IRC images.
The first kind is that seen only in the periphery of the MIR array detectors.
It is thought to originate from the scattering at the detector edge.
An extensive study of the MIR-S array has shown that it is very
stable and thus correctable \citep{Sakon07}.
  
Due to the short baffle of the telescope tube, it was predicted that
the light scattered off the top of the reflective baffle could come into the
telescope at particular seasons of the observation.  This scattered light is
recognized in some IRC images.  The effect of this scattered light on
the flatfields is carefully examined by selecting less-affected data 
and minimized.
This scattered light does not affect observations
of point sources significantly.  
However, observations of diffuse background light
are severely affected and need a special 
caution.

The most serious problem in the IRC calibration at present is the
internal scattering light inside the MIR-L channel.  It is largely noticeable
at L24.  
It affects the flatfield of the MIR-L channels to a level of
10\% as described above.  Ghost images, which might also be related to
this scattered light, are seen against very bright mid-infrared sources.
Characterization of this component is now under investigation. 

\subsection{All-sky survey observation}\label{all-sky}
Two wide bands of S9W and L18W
are used for the IRC mid-infrared all-sky survey observations.
Only the data of two rows of the MIR-S and MIR-L arrays are taken 
in the all-sky survey mode and signals of 4 pixels
in the cross-scan direction are binned on board and transmitted to the
ground  \citep{Ishihara06a}.  
Reading two rows helps in rejecting spurious signals and
high-energy ionization particle hits, and thus significantly
increases the reliability of
source detection.  The virtual pixel scale is about $10^{\prime\prime}
\times 10^{\prime\prime}$ for both bands.  The two rows are
read in an interlaced manner to increase the spatial information.

The absolute calibration 
of the all-sky survey observation is carried out similarly on the basis
of observations of stars of the same standard star network
on orbit \citep{Ishihara07}.  At present, the accuracy in
the absolute calibration is estimated to be 7 and 15\% for S9W and L18W,
respectively.
The 5-$\sigma$ sensitivity for point source detection in a scan
is confirmed to be about 50\,mJy at S9W and 120\,mJy at L18W.  
In the all-sky survey, the sensitivity is limited by detector noise
in both bands.  
These numbers are well
in agreement with the pre-flight expectations as well as the
in-orbit sensitivity for pointed observations.
The IRC mid-infrared all-sky survey observation achieves a deeper
sensitivity and a finer spatial resolution by about an order of
magnitude than the {\it IRAS} 12 and 25\,$\mu$m survey.  Potential
capabilities of the IRC all-sky survey to extragalactic sciences
are discussed in \citet{Pearson04} and \citet{Pearson07}.

\section{Summary}

IRC is one of the two instruments on {\it AKARI} satellite.
It has an imaging capability in the spectral range 1.8 to 26.5\,$\mu$m
with 9 photometric bands and makes wide-field deep imaging in pointed
observations.  It also has a low-resolution spectroscopic
capability, which allows slit-less multi-object spectroscopy in the near
to mid-infrared.  IRC is being used also in an all-sky survey observation mode
at 9 and 18\,$\mu$m with a higher sensitivity and better spatial resolution
by an order of magnitude than the {\it IRAS} 12 and 25\,$\mu$m survey.
It has been confirmed that the in-orbit performance of IRC is well within the
pre-launch predictions and that IRC will provide
significant data in various fields of cutting-edge astronomy.  
IRC data are now being processed with the pipeline software that
takes account of the basic calibrations.  Details are described in
the IRC Data User's Manual \citep{IDUM}.
Examples of IRC data are shown in this issue, which indicate 
the great potential of IRC observations.
The NIR channel is planned to continue observations
after the exhaustion of liquid helium.

\vspace{5pt}
{\it AKARI} is a JAXA project with the participation of ESA.
We thank all the members of the {\it AKARI} project for their continuous 
help and support.  
We thank T. Negishi, I. Maeda, H. Mochizuki, K.-W. Chan,
S. Fujita, and C. Ihara for their 
contributions in the development of the IRC and D. Jennings for providing us
with the on-board calibration sources.  
We also thank K. Imai, M. Ishigaki, H. Matsumoto, N. Matsumoto, and T. Tange
for supporting daily operation of the IRC.
The contribution from the ESAC to the pointing reconstruction 
is greatly 
acknowledged.
W.K., I.S., and T.T. have been financially supported by the
Japan Society of Promotion of Science (JSPS).  This work is supported in part
by a Grant-in-Aid for Scientific Research on Priority Areas from the
Ministry of Education, Culture, Sports, Science, and Technology of Japan
and Grants-in-Aid for Scientific Research from the JSPS.


\end{document}